\documentclass[conference]{IEEEtran}
\IEEEoverridecommandlockouts
\usepackage{cite}
\usepackage{amsmath,amssymb,amsfonts}
\usepackage{graphicx}
\usepackage{textcomp}
\usepackage{xcolor}
\usepackage{multirow}
\usepackage{algorithm,algpseudocode}
\usepackage{algorithmicx}
\usepackage{amsmath}
\usepackage{url}

\def\BibTeX{{\rm B\kern-.05em{\sc i\kern-.025em b}\kern-.08em
    T\kern-.1667em\lower.7ex\hbox{E}\kern-.125emX}}
\begin{document}

\title{A Graphical Method for Identifying Gene Clusters from RNA Sequencing Data}

\author{\IEEEauthorblockN{Jake R. Patock}
\IEEEauthorblockA{\textit{Data to Knowledge Lab} \\
\textit{Rice University}\\
Houston, USA \\
jp157@rice.edu}
\and
\IEEEauthorblockN{Rinki Ratnapriya}
\IEEEauthorblockA{\textit{Department of Ophthalmology} \\
\textit{Baylor College of Medicine}\\
Houston, USA \\
rpriya@bcm.edu}
\and
\IEEEauthorblockN{Arko Barman}
\IEEEauthorblockA{\textit{Data to Knowledge Lab}}
\textit{Rice University}\\
Houston, USA \\
arko.barman@rice.edu}


\maketitle

\begin{abstract}
The identification of disease-gene associations is instrumental in understanding the mechanisms of diseases and developing novel treatments. Besides identifying genes from RNA-Seq datasets, it is often necessary to identify gene clusters that have relationships with a disease. In this work, we propose a graph-based method for using an RNA-Seq dataset with known genes related to a disease and perform a robust clustering analysis to identify clusters of genes. Our method involves the construction of a gene co-expression network, followed by the computation of gene embeddings leveraging Node2Vec+, an algorithm applying weighted biased random walks and skipgram with negative sampling to compute node embeddings from undirected graphs with weighted edges. Finally, we perform spectral clustering to identify clusters of genes. All processes in our entire method are jointly optimized for stability, robustness, and optimality by applying Tree-structured Parzen Estimator. Our method was applied to an RNA-Seq dataset of known genes that have associations with Age-related Macular Degeneration (AMD). We also performed tests to validate and verify the robustness and statistical significance of our methods due to the stochastic nature of the involved processes. Our results show that our method is capable of generating consistent and robust clustering results. Our method can be seamlessly applied to other RNA-Seq datasets due to our process of joint optimization, ensuring the stability and optimality of the several steps in our method, including the construction of a gene co-expression network, computation of gene embeddings, and clustering of genes. Our work will aid in the discovery of natural structures in the RNA-Seq data, and understanding gene regulation and gene functions not just for AMD but for any disease in general.
\end{abstract}

\begin{IEEEkeywords}
RNA-Seq, clustering, transcriptomics, gene co-expression networks, gene embeddings 
\end{IEEEkeywords}

\section{Introduction}
\label{intro}
Gene expression regulation has emerged as a key mechanism through which genetic variants mediate disease risk in humans \cite{doi:10.1126/science.aaz1776}. Thus, studying gene expression data has resulted in the discovery of important links between diseases and genes in recent years \cite{ma2025integrating, cookson_mapping_2009}. Technical advancements in recent years have allowed gene expression instrumentation to produce large amounts of data with lower costs on large samples \cite{muzio2021biological}. Utilizing this gene expression data to discover relationships between genetic factors and phenotypic results across samples and populations has led to new insights on specific diseases and biochemical mechanisms \cite{muzio2021biological}. However, given the size of the datasets and propensity for false positives in gene expression studies, using traditional methods like statistical hypothesis testing leads to noisy results, with additional work being needed to empirically validate the conclusions \cite{ata2021recent}. Given the current difficulties and the potential for translating gene expression findings into precision medicine, improvements in gene expression studies necessitate increased robustness \cite{wellcome2007genome-wide_2007}. 

Many of these difficulties pertaining to identifying the gene expression changes between controls and disease have been shown to be solvable by machine learning (ML) \cite{muzio2021biological}. This potential, combined with increased computational resources, has allowed both classical ML and deep learning (DL) models to be used as effective tools for identifying relevant genes for diseases \cite{muzio2021biological}. The ability of these models to capture complex interdependencies and abstract understandings (in the case of DL) has shown application in multiple gene expression studies in a range of populations \cite{alharbi2022review, ma2025integrating}. Model architectures employed include classic models such as linear regression, logistic regression, and random forest \cite{ma2025integrating} and DL methods, such as graph-based neural networks (GNNs) and graph embedding techniques \cite{ata2021recent}. In particular, DL has the capability to learn interdependencies from genomic expression data and exhibit excellent performance \cite{alharbi2022review}.

A specific application area for utilizing DL in a gene expression application is in the field of Age-related Macular Degeneration (AMD), a neurodegenerative disease that can result in incurable blindness in patients  \cite{ratnapriya2019retinal}. It is a multifactorial disease caused by the cumulative impact of genetic factors, environmental factors, and advanced aging \cite{https://doi.org/10.1111/cge.12206, gorman_genome-wide_2024}. While some genes related to AMD have already been discovered, a large proportion of heritability remains unexplained \cite{https://doi.org/10.1111/cge.12206}, suggesting many undiscovered relationships and dependencies between different genes and the disease. Finding new relationships between genes in AMD patients can lead to new drug targets, patient risk testing, and improved diagnosis for many patients  \cite{ma2025integrating}. Moreover, many known genes associated with AMD may contribute to the disease through distinct biological mechanisms. Grouping these known AMD-related genes into functionally similar clusters could help identify shared pathways or processes, thereby enabling the targeting of underlying disease mechanisms rather than focusing on individual genes. This clustering approach could also help identify previously unrecognized genes associated with AMD by uncovering similarities between undiscovered genes and those already known to be involved in the disease \cite{singh_integration_2025, ma2025integrating}.

This study proposes a graph-based method to discover gene clusters among known genes for a disease from RNA sequencing (RNA-Seq) data. We applied our method to a bulk RNA-Seq dataset for AMD. First, we constructed a gene co-expression network and computed the gene embeddings from the network. Finally, the gene embeddings were clustered to produce a clustering of genes. The results of this workflow can be used to categorize already known genes into groups or clusters, possibly pertaining to similar mechanisms, and produce visual clustering results. Our approach of joint optimization of all involved processes ensures the stability, optimality, and robustness of the entire method. 

In particular, we emphasize our novel approach of jointly optimizing all processes in our method, which enables us to generate the most optimal, robust, and stable clustering results as opposed to individually optimizing each step in most prior work. The final cost function of our joint optimization is based on the clustering performance in the overall process, which yields superior performance compared to separately optimizing co-expression network construction, embedding computation, and clustering that would guarantee optimal results for individual steps but not the overall process. 

Our contributions can be summarized as follows:
\begin{itemize}
    \item The design of a novel pipeline for clustering genes from RNA-Seq data, involving the construction of a co-expression network, computing gene embeddings from the network, and clustering genes based on their embeddings.
    \item The joint optimization of all involved steps in the pipeline for the best possible clustering results.
    \item The design of a set of tests to ensure the robustness, consistency, and statistical significance of the results.
\end{itemize}

\section{Related Work}
\label{related_work}

\begin{figure*}[t]
    \centering
    \includegraphics[trim=40 200 230 30, clip,width=.8\textwidth]{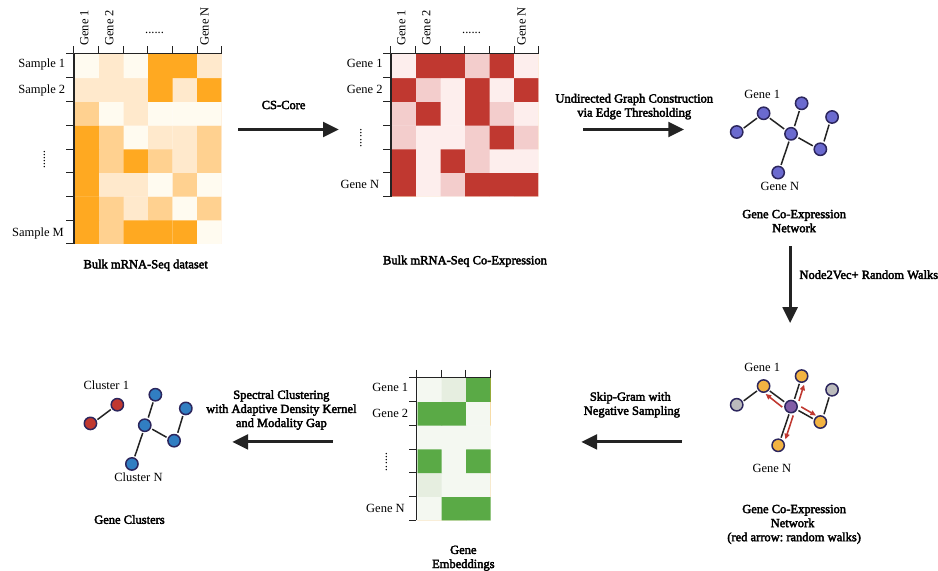}
    \caption{Overview of our method used to process the subset of bulk mRNA-seq dataset comprising 81 AMD genes to generate a co-expression matrix using CS-CORE, generating a co-expression network by thresholding, embedding to a latent space using \texttt{Node2Vec+}, and finally applying spectral clustering.}
    \label{fig:method_overview}
\end{figure*} 

In recent years, ML models have been successfully applied to RNA-Seq datasets for different diseases, such as breast cancer, colorectal cancer, and other types of cancers, and AMD\cite{cheng_machine_2024, cascianelli_machine_2020, taghizadeh_breast_2022, maurya_transcriptome_2021, jha_identifying_2022, ma2025integrating}. Proposed methods and results show success in stratification of breast cancer into molecular subtypes using multiclass logistic regression \cite{cascianelli_machine_2020}, ML-based identification of differentially expressed genes and independently modulated gene sets \cite{wang_rna-seq_2018, rychel_machine_2020}, identification of transcriptomic features that are commonly shared between cancer types \cite{jha_identifying_2022}, identification of diagnostic biomarkers for colorectal cancer \cite{maurya_transcriptome_2021}, breast cancer prediction with transcriptome profiling \cite{taghizadeh_breast_2022}, and prediction of gene regulatory networks \cite{mochida_statistical_2018}. A wide variety of both supervised and unsupervised classical ML algorithms, such as support vector machines (SVM), Extreme Gradient Boosting (XGBoost), self-organizing maps (SOM), different dimensionality reduction algorithms, such as t-distributed Stochastic Neighbor Embedding (tSNE) and PCA, and clustering algorithms, such as k-means and hierarchical clustering, have been used for the analysis of RNA-Seq datasets \cite{cheng_machine_2024, petegrosso_machine_2019}. The use of graph-based DL and ML algorithms, such as knowledge graphs \cite{torgano_rna_2025} and Node2Vec \cite{grover2016node2vec}, for transcriptomics is also being used in recent years for research in different diseases, including melanoma \cite{wang_single-cell_2021}.

\section{Methods}
\subsection{Overview}
The purpose of this study was to perform a gene expression analysis to embed and cluster a group of 81 previously identified genes related to AMD \cite{ma2025integrating}. These genes were identified by applying ML methods on an mRNAseq dataset, which contained the bulk gene expression levels for control and late-stage positive AMD patients. This set of AMD genes was experimentally validated to be statistically significant in separating the controls from late-stage AMD subjects \cite{ma2025integrating}. 

An overview of our methods is shown in Figure \ref{fig:method_overview}. First, we constructed an undirected graph based on the co-expression data calculated from the dataset. To control for varying sequence depth errors, CS-CORE (cell-type-specific co-expressions), a statistical method, was used to derive the co-expression matrix from our subset of 81 genes \cite{su2023cell}. The graph was then constructed between the genes (nodes) with edges being added to the graph where the co-expression coefficient's absolute value is greater than a specific threshold, $\tau$, that can be tuned as a hyperparameter. 

The resulting graph is then used for obtaining node embeddings (i.e., gene embeddings) in a latent space, by applying the \texttt{Node2Vec+} algorithm \cite{10.1093/bioinformatics/btad047}. In \texttt{Node2Vec+}, the first step is to perform biased weighted random walks. The resulting anchor node inputs and neighboring nodes are then used as input and labels for Skip-Gram with Negative Sampling (SGNS) to produce the embedding for each node. These node embeddings were then clustered using spectral clustering, utilizing multimodality gap for optimizing the number of clusters \cite{grover2016node2vec}. 

Finally, the output clusters were evaluated to determine the statistical significance of the clustering of the 81 gene subset versus random samples of 81 genes from the entire set of genes. The entire pipeline, including the generation of the co-expression network, calculation of the gene embeddings in a latent space through \texttt{Node2Vec+}, and spectral clustering, was optimized jointly by applying Tree-structured Parzen Estimator \cite{bergstra_making_2013}. We emphasize that our joint optimization strategy results in more consistent, robust, and optimal clustering results, as all processes are optimized jointly with a single cost function instead of optimizing each step separately with separate cost functions, which cannot guarantee an optimal final solution.

Since \texttt{Node2Vec+} (including biased weighted random walks) and spectral clustering are stochastic processes, we repeated these processes $100$ times, leading to $100$ different embeddings along with corresponding clusterings, to verify and validate the robustness and consistency of our methods. 


\subsection{Dataset}
The dataset used in this work includes the complete mRNAseq data from $n_c=105$ controls (Minnesota Grading System 1) and $n_s=61$ subjects (Minnesota Grading System 4) with late-stage AMD~\cite{ratnapriya2019retinal}. In particular, for applying our entire pipeline, we used only a subset of 81 genes from this dataset that are related to AMD as previously identified using ML methods~\cite{ma2025integrating}. These 81 previously identified genes were validated to be statistically significant, and novel disease-gene associations were identified through the use of feature selection, XGBoost, and SHapley Additive exPlanations (SHAP) to create an interpretable ML pipeline to identify and discover disease-gene associations.

\subsection{Gene Co-expression Network Construction from CS-Core}
To calculate the co-expression levels within the dataset, CS-CORE was applied \cite{su2023cell}. CS-CORE is a statistical method that accounts for varying sequence depth in genetic co-expression calculations within mRNAseq datasets \cite{su2023cell}. This method leads to a more accurate estimation of co-expression between genes compared to other methods, such as the Pearson correlation coefficient \cite{su2023cell}. CS-CORE was shown to have a much better ability to adjust for false positives, more accurate co-expression level estimation, and expression detection in both Alzheimer's disease and COVID-19~\cite{su2023cell}. In our methods, we assumed that CS-CORE is applicable to bulk mRNAseq data in the absence of single-cell mRNAseq data. 

From this co-expression matrix, a gene co-expression network was constructed, where the nodes of the graph were individual genes, with an undirected weighted edge added between a pair of nodes if the CS-CORE co-expression value for the pair was greater than a threshold, $\tau$. This edge threshold hyperparameter,
$\tau$, was tuned in the optimization stage outlined in section \ref{section:hypers}. 

\subsection{Graph Embedding Using Node2Vec+} 
From the gene co-expression network, we calculated node embeddings, i.e., gene embeddings, in a latent space using \texttt{Node2Vec+} \cite{10.1093/bioinformatics/btad047}. \texttt{Node2Vec+} refines the classic \texttt{Node2Vec} algorithm for computing the node embeddings in a graph \cite{grover2016node2vec}. In the first stage of \texttt{Node2Vec}, an anchor node is chosen and random graph traversals are executed to select neighbor nodes that are accessible from the anchor node by performing weighted biased random walks (based on edge weights), and a list of visited nodes is recorded \cite{grover2016node2vec}. These anchor nodes are then used as inputs to a Skip-Gram with Negative Sampling (SGNS) model, with the corresponding neighborhood nodes being the output labels \cite{grover2016node2vec} to finally train the \texttt{Node2Vec} model to obtain node embeddings. 

In \texttt{Node2Vec+}, edge weights are considered on this random walk through the hyperparameters -- the return hyperparameter $p$, the in-out hyperparameter $q$, and a looseness hyperparameter $\gamma$ \cite{10.1093/bioinformatics/btad047}. These hyperparameters can bias the model towards moving down edges with higher weights or returning to previously traversed edges \cite{10.1093/bioinformatics/btad047}. 

Overall, we optimized the following hyperparameters in \texttt{Node2Vec+} -- return hyperparameter ($p$), in-out hyperparameter ($q$), length of each walk from an anchor node ($W_L$), embedding dimensionality ($E$), window size ($W_S$), and the number of negative samples per positive sample ($N_s$). The hyperparameter, $\gamma$, when set to $0$, was shown to be robust and $\gamma=0$ was used in our experiments \cite{10.1093/bioinformatics/btad047} (refer to Section \ref{section:hypers} for more details).

Overall, the refinement introduced by \texttt{Node2Vec+} improves the effectiveness of the second-order random walk of \texttt{Node2Vec} by considering the edge weights of the graph and thus yields better node embeddings, especially for biological networks and datasets \cite{10.1093/bioinformatics/btad047}. Thus, we chose \texttt{Node2Vec+} over the classic \texttt{Node2Vec} to embed the genes in the co-expression network into a latent node embedding space for the AMD gene dataset. 

Three hyperparameters related to SGNS were optimized -- the embedding dimension ($E$), window size ($W_S$), and number of negative sampling predictions ($N_S$) during the training of the SGNS. For fitting the SGNS model, it was trained for $100$ epochs, and $32,768$ random walks were performed on the graph to generate the dataset for each model trial.

\subsection{Spectral Clustering}
Spectrum, a spectral clustering method for multi-omic datasets, was applied to the node embeddings computed using \texttt{Node2Vec+}. To calculate the affinity matrix of the embeddings, we applied the adaptive density-aware kernel designed in Spectrum \cite{john2020spectrum}. The adaptive density-aware kernel defines the affinity between two nodes, $s_i$ and $s_j$, as,
\begin{equation}
A_{ij} = \exp\left(-\frac{d^2(s_i,s_j)}{\sigma_i \sigma_j(CNN(s_i s_j) + 1}\right)
\end{equation}
where
\begin{itemize}
    \item $d(s_i, s_j) \in \mathbb{R}$ is the Euclidean distance between nodes $s_i$ and $s_j$.
    \item $\sigma_i \in \mathbb{R}$ is the local scale parameter for the node $s_i$, calculated as the Euclidean distance between node $s_i$ and its $P$-th nearest neighbor, $s_{i,P}$, i.e., $\sigma_i=d(s_i, s_{i,P})$,
    \item $\sigma_j \in \mathbb{R}$ is the local scale parameter for the node $s_j$, calculated as the Euclidean distance between node $s_j$ and its $P$-th nearest neighbor, $s_{j,P}$, i.e., $\sigma_i=d(s_j, s_{j,P})$, and
    \item $CNN(s_i s_j)$ is the number of points in the intersection of the sets of $S$ nearest neighbors of $s_i$ and $s_j$.
\end{itemize}

The hyperparameters $P$ and $S$ were jointly optimized along with the other hyperparameters in our entire process. This method was found to enhance the similarities between two nodes in a graph if they shared similar connectivity. 

To estimate the number of clusters, the last substantial multimodality gap heuristic was employed~\cite{john2020spectrum}. First a diagonal matrix, $D$, was constructed, where $D_{ii} = \sum_j A_{ij}$. The normalized graph Laplacian was then computed as,
\begin{equation}
    L = D^{-1/2} A D^{-1/2}
\end{equation}
We performed the eigendecomposition of $L$ to obtain eigenvectors, $V = \{x_1, x_2, \dots, x_N\}$, and eigenvalues, $\Lambda = \{\lambda_1, \lambda_2, \dots, \lambda_N\}$. Multimodality was quantified by calculating the dip test statistics, $Z = \{z_1, z_2, \dots, z_N\}$, for all eigenvectors. Finally, multimodality gaps were computed as $d_i = z_i - z_{i-1}$. The last substantial multimodality gap was calculated by iterating through $d_i$ from the largest to the smallest corresponding eigenvalue using a search method that yields the optimal number of clusters, $k^*$~\cite{john2020spectrum}. 

For clustering, the top $k^*$ eigenvectors were stacked to form a matrix, $X = [x_1 \; x_2 \; \dots \; x_{k^*}] \in \mathbb{R}^{N \times k^*}$. A row-normalized matrix, $Y \in \mathbb{R}^{N \times k^*}$ was constructed, where,
\begin{align}
    Y_{ij} = \frac{X_{ij}}{\left(\sum_j X_{ij}^2\right)^{1/2}}
\end{align}

A Gaussian Mixture Model (GMM) was used to obtain $k^*$ gene clusters, considering each row of $Y$ as a data point. 

\subsection{Hyperparameter Optimization}
\label{section:hypers}
All the steps in our methods -- construction of the gene co-expression network from CS-CORE, Node2Vec+, and spectral clustering -- involve hyperparameters that need to be tuned. Instead of optimizing each process individually, which might not result in the optimal clustering results, we chose to jointly optimize all the hyperparameters in our methods to maximize the density-based clustering validation index (DBCVI) \cite{moulavi2014density}. DBCVI is suitable for non-convex clustering algorithms, such as spectral clustering, and has been shown to be effective for various applications and datasets \cite{moulavi2014density}. 

The following processes and hyperparameters were jointly optimized in our experiments:
\begin{itemize}
    \item Construction of the gene co-expression network from CS-CORE: 
    \begin{itemize}
        \item threshold of edge weights for connectivity between nodes, $\tau$
    \end{itemize}
    \item Computation of gene embeddings using \texttt{Node2Vec+}:
    \begin{itemize}
        \item return hyperparameter, $p$
        \item in-out hyperparameter, $q$
        \item walk length, $W_L$
        \item embedding dimensionality, $E$
        \item window size, $W_S$
        \item number of negative samples per positive sample, $N_s$
    \end{itemize}
    \item Spectral Clustering:
    \begin{itemize}
        \item nearest neighbor hyperparameter for calculating local scale parameters, $P$
        \item number of nearest neighbors for calculating intersection, $S$
    \end{itemize}
\end{itemize}

The complete search space for the hyperparameters is outlined in \ref{tab:search_space}. Tree-structured Parzen Estimator (TPE), a Bayesian optimization method, was used to optimize this set of hyperparameters using 256 hyperparameter samplings from the search space. Since the process of random walks in \texttt{Node2Vec+} is stochastic, every hyperparameter configuration was tried 8 times and averaged to produce the final DBCVI output, to average the effects of stochasticity on the results. 


Our algorithm, including co-expression network construction, \texttt{Node2Vec+} for embedding calculation, and spectral clustering, along with our joint hyperparameter optimization strategy, is shown in Algorithm \ref{algo:our_algo}.

\begin{algorithm}[tbh]
\caption{Our Algorithm}
\label{algo:our_algo}
\begin{algorithmic}[1]
\Require Bulk mRNA-seq dataset $X \in \mathbb{R}^{n \times g}$, hyperparameter search space $\mathcal{H}$, trials $N$
\Ensure Optimal hyperparameter configuration for clustering score
\State \textbf{Initialize:} TPE sampler; objective: maximize DBCVI
\For{trial $t = 1$ to $N$}
    \State Sample hyperparameters from TPE: 
    \Statex \hspace{\algorithmicindent} $\{\tau, p, q, W_L, E, W_S, N_S, P, S\} \sim \text{TPE}(\mathcal{H})$
    \State Initialize $\text{DBCVI\_Scores} \leftarrow [ \, ]$
    \For{repeat $r = 1$ to $8$}
        \State $\mathbf{C} \leftarrow \text{CS-CORE}(X)$
        \State $G \leftarrow \text{ConstructGraph}(\mathbf{C}, \tau)$
        \State $\mathbf{Z} \leftarrow \texttt{Node2Vec+}(G, p, q, W_L, E, W_S, N_S)$
        \State $\text{Clusters} \leftarrow \text{SpectralClustering}(\mathbf{Z}, P, S)$
        \State $\text{dbcvi} \leftarrow \text{DBCVI}(\text{Clusters}, \mathbf{Z})$
        \State Append $\text{dbcvi}$ to $\text{DBCVI\_scores}$
    \EndFor
    \State $\overline{\text{DBCVI}}_t \leftarrow \frac{1}{8} \sum_{r=1}^{8} \text{DBCVI\_scores}[r]$
    \State Store $(\{\tau, p, q, W_L, E, W_S, N_h, S, P\}, \overline{\text{DBCVI}}_t)$
\EndFor
\State \textbf{Return:} $\mathcal{H}^* \leftarrow \arg\max_{t \in [1,N]} \overline{\text{DBCVI}}_t$
\end{algorithmic}
\end{algorithm}


\begin{table}[h]
\centering
\caption{Hyperparameter search space optimized via TPE to maximize DBCVI of output clusterings, with corresponding optimal values.}
\begin{tabular}{|l|l|c|c|}
\hline
\textbf{Method} & \textbf{Hyperparameter} & \textbf{Search Space $\mathcal{H}$} & \textbf{Optimum} \\
\hline
\multirow{1}{*}{Graph Construction} 
  & \texttt{$\tau$}   & [0.3, 0.5]       & 0.4 \\
\hline
\multirow{6}{*}{Node2vec+} 
  & \texttt{p}                & [0.01, 100.0]    & 0.35  \\
  & \texttt{q}                & [0.01, 100.0]    & 11.66 \\
  & \texttt{$W_L$}      & [10, 30]         & 20 \\
  & \texttt{$E$}   & [2, 16]          & 10 \\
  & \texttt{$W_S$}      & [4, 10]          & 10 \\
  & \texttt{$N_s$} & [5, 15]          & 7 \\
\hline
\multirow{2}{*}{Spectral Clustering} 
  & \texttt{P}                & [3, 7]           & 7 \\
  & \texttt{S}                & [P+2, P+4]       & 11 \\
\hline
\end{tabular}
\label{tab:search_space}
\end{table}


\subsection{Testing Statistical Significance and Robustness}
We performed experiments to validate the robustness and statistical significance of our clusterings for the AMD gene dataset. We show that our clusterings are significantly different from clusterings of 81 genes randomly sampled from the entire set of genes in the data. For this experiment, we generated 100 randomly sampled sets of 81 genes. Our methods were applied to these randomly-sampled sets -- CS-CORE, co-expression network construction, \texttt{Node2Vec+}, and spectral clustering. Finally, to evaluate the clustering results, DBCVI was calculated for each of these 100 randomly-sampled sets. The distribution of DBCVI values from these randomly-sampled sets was then compared with the distribution of DBCVI values from 100 repetitions of clustering the known 81-gene AMD dataset using statistical tests. In particular, the Kolmogorov–Smirnov (K-S) test and the k-sample Anderson-Darling test were applied to test if the differences between these distributions were statistically significant.

Due to the inherent stochasticity in \texttt{Node2Vec+} and spectral clustering, it is imperative to test the robustness of our methods to ensure that the results are consistent to account for stochasticity. To test the robustness of our methods, we performed 100 repetitions of our entire method of clustering. Every pair of clustering results in these 100 repetitions was compared with each other using the adjusted mutual information (AMI) score to evaluate their agreement \cite{nguyen2009information}. AMI has been shown to be the ideal metric in the presence of small clusters and unbalanced cluster sizes. The range of values for AMI is -1 to 1.


\section{Results}
Joint hyperparameter optimization was performed for all hyperparameters using 256 samplings from the hyperparameter space to maximize DBCVI for the clustering results. The optimal hyperparameters are shown in Table \ref{tab:search_space}. This optimal set of hyperparameters yielded a DBCVI score of 0.99 over the 8 clustering trials per hyperparameter configuration. We note that the optimal gene embedding dimensionality was 10.

\begin{figure}[t]
    \centering
    \includegraphics[trim=70 70 140 180, clip, width=\linewidth]{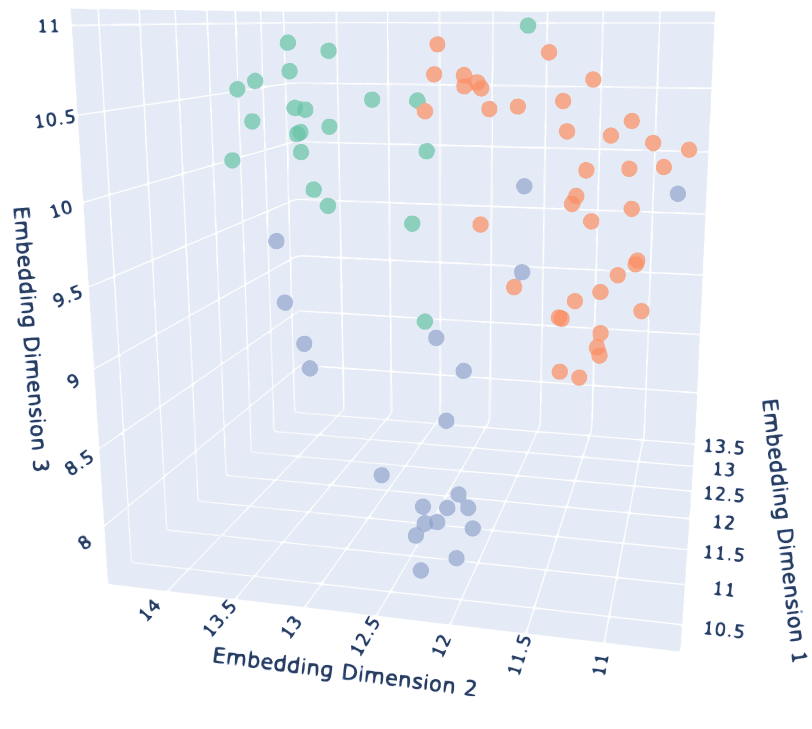}
    \caption{Clustering results for the known 81-gene AMD subset. Embedding dimensionality is reduced from 10 to 3 using UMAP for visualization. (An interactive HTML visualization is available in our code repository.)}
    \label{fig:example_clustering}
\end{figure}

After optimization, this optimal set of hyperparameters was then used to embed and cluster the dataset 100 times to perform statistical significance and robustness tests. The average DBCVI score for these 100 repetitions was 0.95, showing that each trial resulted in well-formed clusters. In our robustness testing, the average AMI between all pairs of clustering results was calculated to be $0.49$, with a variance of $0.022$. Hence, the 100 clustering repetitions show moderately to highly consistent agreement in the clustering of genes, especially for a small dataset with the possible presence of noise.

For visualization, the dimensional gene embeddings (of optimal dimensionality 10) were reduced to 3 dimensions using Uniform Manifold Approximation and Projection (UMAP) \cite{mcinnes2020umapuniformmanifoldapproximation}. These low-dimensional embeddings were then plotted to produce interpretable results for AMD medical professionals to utilize in their research. The clustering visualization is shown in Figure \ref{fig:example_clustering}.

\begin{figure}[t]
    \centering
    \includegraphics[width=\linewidth]{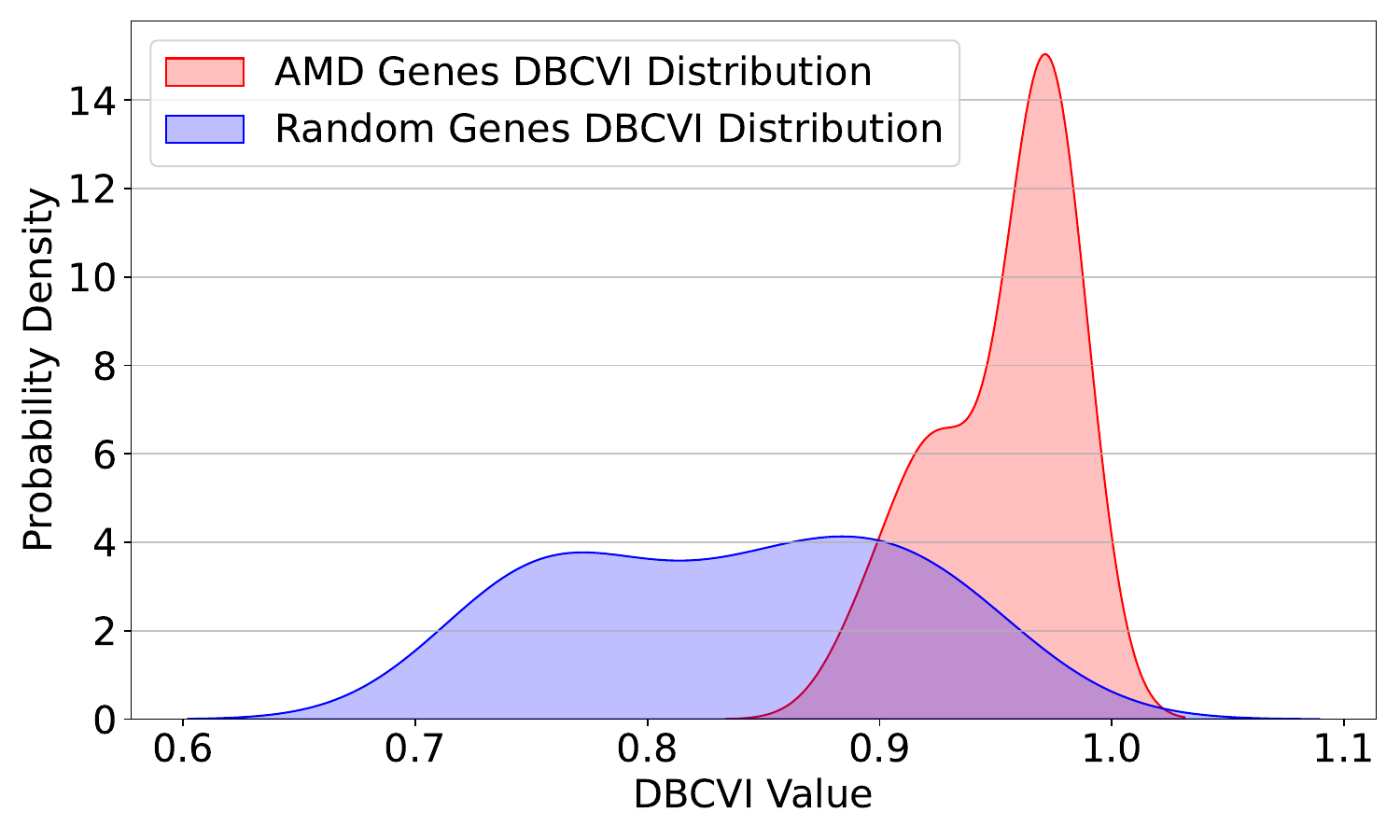}
    \caption{Distributions of DBCVI values for 100 clustering repetitions for the known 81-gene subset for AMD and 100 samples of 81 randomly sampled genes.}
    \label{fig:kde}
\end{figure}

In our statistical significance testing, the distribution of the DBCVI values for the 81-gene known subset of AMD genes was compared with that of 100 randomly sampled subsets of 81 genes from the entire set of genes. While the DBCVI for the known 81-gene subset yielded an average DBCVI of $0.95$, the average DBCVI for the randomly sampled genes was $0.84$. These distributions of the DBCVI values are shown in Figure \ref{fig:kde}. K-S test performed to validate that the differences between these distributions are statistically significant yielded $p=2.68 \times 10^{-25}$. We also applied the k-sample Anderson-Darling test, a non-parametric test, to check if our clustering results were significantly different from the clusterings of the randomly sampled gene subsets, which yielded $p<0.001$. Thus, our methods, when applied to the known 81-gene subset, yield significantly better clustering results than a randomly selected subset of 81 genes.

\section{Discussion}
Our method of embedding and clustering the bulk mRNAseq dataset showed robust, statistically significant results. The AMI score indicated a moderate to high level of agreement between clustering results. Considering the high level of noise typically present in transcriptomic datasets, the AMI scores show the robustness of our methods. We note that we employed our methods on a dataset of limited sample size, and using larger datasets is expected to lead to more consistent results and higher AMI.

We also noted that a higher number of random walks results in higher AMI. In the presence of more computing availability, the number of walks could continue to be scaled up to increase AMI further. Moreover, increasing the random walks performed on the graph will allow larger datasets as inputs to SGNS, leading to higher AMI. This hyperparameter can be dynamically adjusted based on the dataset on which this method is used. 

The use of CS-CORE instead of Pearson correlation generated a more refined construction of gene co-expression network, resulting in better gene embeddings with the application of \texttt{Node2Vec+}, which in turn, resulted in more robust clustering results. Finally, spectral clustering leveraging the adaptive density kernel with modular gap metric showed robustness in consistently clustering these gene embeddings. Finding consistent clustering has been a long-standing challenge in various transcriptomics methods. Our method showed promise in providing a solution for our dataset and can possibly be extended to other transcriptomic analyses and datasets.

Our method is highly versatile due to the joint optimization of hyperparameters for all the steps involved in the entire process and can easily be applied to other datasets. This versatility and robustness allow our model to be applied to other datasets and diseases for discovering gene clusterings. Further, our method is computationally feasible without the use of GPUs or other computationally expensive hardware, making it widely accessible to researchers without any barriers related to computational capabilities and the availability of specialized hardware. Overall, our method is flexible, accessible, scalable, and generalizable.

Employing our method on more diverse mRNAseq datasets would be helpful for future research on other diseases. Our method can be applied to any available mRNAseq dataset. For future work, testing our method on a synthetic dataset with known ground-truth labels would allow for a more rigorous evaluation of its ability to recover the true information structure within the data. This approach would validate the results independently, complementing our experiments to test robustness and statistical significance. 

\section{Conclusion}
In this work, we proposed a comprehensive pipeline for discovering gene clusters by constructing a gene co-expression network, computing gene embeddings, and finally clustering genes in an mRNA-seq dataset for AMD. Our approach integrates multiple stages to ensure robust results. First, we apply CS-CORE to correct for expression errors and normalize the data. Next, we construct a gene co-expression network and calculate gene embeddings by applying \texttt{Node2Vec+}. Finally, we perform spectral clustering to cluster genes and identify gene clusters that have a potential relationship with the disease.

The workflow demonstrates high adaptability, as it can be tuned to the specific characteristics of any given dataset through joint hyperparameter optimization and consistently identifies cluster patterns across multiple repetitions, even though it has several stochastic components. Our results indicate that our method is flexible and robust in capturing meaningful biological structure in transcriptomic data. Notably, our method identifies gene clusters that experimental molecular geneticists can directly validate, interpret, and leverage for future research. These clusters can inform experimental design, guide hypothesis design, and be incorporated into existing laboratory workflows, ultimately supporting decision-making in studies of gene function and disease mechanisms.

Our code and results are available at \url{https://github.com/arkobarman/graphical_gene_clustering_RNASeq}.

\bibliographystyle{plain} 
\bibliography{main}

\end{document}